\documentclass[prl,aps,twocolumn]{revtex4}
\usepackage{graphicx}
\usepackage{amsmath}
\usepackage{amssymb}
\usepackage{bm}

\newcommand{\be}{\begin{equation}}
\newcommand{\ee}{\end{equation}}
\newcommand{\bea}{\begin{eqnarray}}
\newcommand{\eea}{\end{eqnarray}}
\newcommand{\beas}{\begin{eqnarray*}}
\newcommand{\eeas}{\end{eqnarray*}}
\newcommand{\bei}{\begin{itemize}}
\newcommand{\eei}{\end{itemize}}

\def\unue#1{{\it#1:}}

\begin{document}

\title{Designing Topological Bands in Reciprocal Space}

\author{N. R. Cooper$^1$ and R. Moessner$^2$}
\affiliation{\parbox{14cm}{$^1$Cavendish Laboratory,~University of Cambridge,~J.~J.~Thomson~Ave.,~Cambridge CB3~0HE,~U.K.}\\
\parbox{14cm}{
$^2$Max-Planck-Institut f{\" u}r Physik komplexer Systeme, N{\" o}thnitzer Stra\ss e 38, Dresden, Germany}}

\begin{abstract}

  Motivated by new capabilities to realise artificial gauge fields in
  ultracold atomic systems, and by their potential to access
  correlated topological phases in lattice systems, we present a new
  strategy for designing topologically non-trivial band
  structures. Our approach is simple and direct: it amounts to
  considering tight-binding models {\it directly in reciprocal
    space}. These models naturally cause atoms to experience highly
  uniform magnetic flux density and lead to topological bands with
  very narrow dispersion, without fine-tuning of parameters. Further,
  our construction immediately yields instances of optical Chern
  lattices, as well as band structures of higher Chern number, $|{\cal
    C}|>1$.

\end{abstract} 
\date{\today}



\maketitle

The ability to create new environments for custom-tailored degrees of
freedom is central to advancing our understanding of correlated
electron physics. In the same way that modulation-doped semiconductor
heterostructures gave rise to the field of fractional quantum Hall
physics, the creation of artifical gauge fields\cite{dalibardreview}
holds the promise of providing an entirely new perspective on
strong-correlation physics in topological
bands. 
The question that naturally arises now is not only how to generate
such artificial gauge fields, but also what settings will be most
promising for realising new physical phenomena.
There is intense current interest in the properties of such systems in
condensed matter physics, most recently through the proposed existence
of fractional Chern insulators\cite{roysondhisynopsis,
  PhysRevLett.106.236802,PhysRevLett.106.236803,PhysRevLett.106.236804}. These
are lattice systems in which fractional quantum Hall physics occurs in
partially filled non-dispersive topological ``Chern''
bands\cite{thoulesschern,haldanehoneycomb}. 
These works are all based on tight-binding models, in which
band topology and dispersion are controlled by the tunneling matrix
elements.
The first steps have been taken in implementing 
tight-binding models of this type for atoms in optical
lattices\cite{lewensteinrev}, with tunable tunneling
phases\cite{JakschZoller} at least for nearest-neighbour
sites\cite{Aidelsburger,Jimenez,Struck}.

Recently it has been demonstrated that topological bands can be formed
for ultracold atoms in another, very direct,
way\cite{ofl,bericooperz2,cooperdalibard}. Coherent Raman coupling of several ($N>1$)
internal atomic states\cite{dalibardreview}, as used to generate
artificial gauge fields\cite{spielmanfield},
 allows new forms of
optical lattice.  In particular, in ``optical flux lattices''  atoms
experience non-zero average magnetic flux density, resulting in
low-energy topological bands that are analogous to Landau
levels\cite{ofl,cooperdalibard}. These new forms of optical lattice
are readily implemented in experiment, so it is highly desirable to be
able to design their properties.  However, they can operate far from the
tight-binding limit, so the existing solid-state models cannot be
directly applied.

Here, we present a number of insights into the nature of
optical flux lattices and develop theoretical tools that allow a
targeted design of topological bands in optical lattices.  Notably, we
show that optical flux lattices are usefully viewed as
tight-binding models {\it in reciprocal space}.  This allows a design
of band structures by transposing established results on Chern bands
in real space into our new setting. We use our insights to construct
three new forms of optical lattice.  First, we show how to generate
optical flux lattices, involving more than $N=2$ internal states, for
which the  magnetic flux density is highly uniform in
space. The resulting energy bands have very narrow bandwidth, and
recover continuum Landau levels for large $N$.  Second, we show how to
construct ``optical Chern lattices'', in which the dressed states
experience zero net magnetic flux, but for which the band has
non-zero Chern number. Finally, we show how one can design optical
lattices in which the lowest energy band has Chern number larger than
one.

Our results follow from examining the limits of deep/shallow
lattices, when the lattice depth ${\cal V}$ is large/small
compared to the typical kinetic energy $E_{\rm R}\equiv
\hbar^2\kappa^2/(2M)$ (all symbols are defined below).  
After setting up the problem, we discuss these two limits in turn. We
comment on possible implementations on the way.

\label{sec:tb}

The Hamiltonian for an optically dressed atom with $N$ internal states
is\cite{dalibardreview}
\be \hat{H} =
\frac{{{\bm p}}^2}{2M} \hat{\openone}_N + \hat{V}({\bm
  r}) \label{eq:ham} \ee
where $\hat{\openone}_N$ is the $N\times N$ identity matrix, and
$\hat{V}({\bm r})$ an $N\times N$ matrix describing the coherent
optical coupling in the rotating wave approximation.  The nature of
the internal states is unimportant for our purposes: these can be
electronic excited states, hyperfine states, or states of orbital
motion.

In all cases, the coupling $\hat{V}({\bm r})$  involves
absorption/emission of photons from laser beams, so it naturally
appears as couplings $V^{\alpha'\alpha}_{{\bm k}'-{\bm k}}
\equiv \langle \alpha', {\bm k}' |\hat{V} |\alpha, {\bm k}\rangle$,
where $|\alpha,{\bm k}\rangle $ is the momentum ${\bm k}$ eigenstate
for component $\alpha$.
We focus on periodic lattices, such that the set of momentum transfers
$\{{\bm \kappa}\}$ for which $V^{\alpha'\alpha}_{{\bm \kappa}}$ is
non-zero forms a regular lattice. The momentum ${\bm k}$ of any given
component $\alpha$ is then only conserved up to the addition of
reciprocal lattice vectors, the basis vectors of which we denote ${\bm
  G}_i$, with $i=1\ldots d$ where $d$ is the dimensionality of the
lattice. (We focus on lattices in $d=2$, but the results can be
extended to $d=3$, with Berry curvature and magnetic flux density
becoming pseudovectors.)

By Bloch's theorem, the energy eigenstates of (\ref{eq:ham}) can be assigned
crystal momentum ${\bm q}$  and band index $n$, and 
be decomposed as
$
|\psi^{n{\bm q}}\rangle = \sum_{\alpha,{\bm G}} c^{n {\bm q}}_{\alpha {\bm G}} |\alpha, {\bm q} - {\bm g}_\alpha -{\bm G}\rangle
$
where $\alpha$ runs over all $N$ internal states,
${\bm G}$ runs over all
sites of the reciprocal lattice,
and the vectors
${\bm g}_\alpha$ account for possible momentum offsets involved in the
inter-state transitions.
 The energy eigenvalues $E_{n{\bm q}}$
follow from
\bea
E_{n{\bm q}} c^{n {\bm q}}_{\alpha {\bm G}}  & = & 
\epsilon_{{\bm q} - {\bm g}_\alpha -{\bm G}} c^{n {\bm q}}_{\alpha {\bm G}} 
+\sum_{\alpha',{\bm G}'}   V^{\alpha'\alpha}_{ {\bm g}_{\alpha} + {\bm G} - {\bm g}_{\alpha'} -{\bm G}'}
 c^{n {\bm q}}_{\alpha' {\bm G}'}
\label{eq:tb}
\eea 
where $\epsilon_{\bm k} \equiv \hbar^2|{\bm k}|^2/(2M)$.

\unue{Deep lattice/adiabatic limit} In the limit $E_{\rm
    R}/{\cal V}\to 0$ the kinetic energy can be neglected. The
effects of the optical coupling in Eqn.~(\ref{eq:ham}) can then
be fully understood in terms of the local dressed states, the
eigenstates of $\hat{V}({\bm r})$.  Under adiabatic motion, the $n^{\rm th}$ dressed state
experiences magnetic flux density\cite{dalibardreview}
$n_{\phi}({\bm r}) = \frac{\hat{\bm z}\cdot}{2\pi i} \sum_\alpha {\bm \nabla}u^{n*}_\alpha \times {\bm \nabla}u^n_\alpha$,
due to the Berry curvature\cite{niureview} associated with spatial
variations of its wavefunction $u^n_\alpha({\bm r})$.  In an optical
flux lattice, the lowest energy dressed state experiences $N_\phi\neq
0$ magnetic flux quanta through each unit cell\cite{ofl}.

It is instructive to examine this limit $E_{\rm R}/ {\cal
    V}\to 0$ also from the point of view of Eqn.(\ref{eq:tb}). Dropping the
  kinetic energy causes this eigenvalue problem to reduce to that of
a {\it uniform tight-binding model in reciprocal space}, defined by
the couplings $V^{\alpha'\alpha}_{ {\bm g}_{\alpha} + {\bm G} - {\bm
    g}_{\alpha'} -{\bm G}'}$ between sites at positions ${\bm
  g}_\alpha+{\bm G}$.  The energy spectrum of this reciprocal-space
tight-binding model -- its ``bandstructure'' -- consists of $N$ bands,
since there are $N$ sites ${\bm g}_\alpha$ associated with each
reciprocal lattice point ${\bm G}$.  The eigenstates are extended
Bloch waves
in reciprocal space, and can be written $c^{n {\bm r}}_{\alpha {\bm
    G}} \propto e^{i({\bm G}+{\bm g}_\alpha)\cdot {\bm
    r}}u^n_\alpha({\bm r})$. (For $E_{\rm R}/{\cal V}\to 0$ the
energy is independent of ${\bm q}$ and the states are naturally labelled
by the band index $n=1\ldots N$ and a conserved ``momentum''
${\bm r}$.)

Crucially, we identify this conserved ``momentum'' of the
reciprocal-space tight-binding model with the real space position
${\bm r}$ (and thus its ``Brillouin zone'' with the real space unit
cell). The state in the $n^{\rm th}$ band simply corresponds to the
$n^{\rm th}$ local dressed state of $\hat{V}({\bm r})$.  The power of
this viewpoint emerges upon considering the magnetic flux density
$n_\phi$: the Berry curvature associated with the adiabatic motion of
any dressed state in real space equals the Berry curvature of the
associated band of the reciprocal-space tight-binding model.  In
particular, the number $N_\phi$ of magnetic flux quanta through the
real space unit cell equals the Chern number of that band.  Hence,
{\it the criterion for an optical flux lattice is that the lowest
  energy band of the reciprocal-space tight-binding model has non-zero
  Chern number.}  Since much is known about the Chern bands of various
tight-binding models, this criterion can be used to design optical
flux lattices with specific properties.

One highly desirable feature is to generate a magnetic flux density
that is uniform. The resulting energy bands will then closely
reproduce Landau levels: topological bands with exact degeneracy,
highly susceptible to strong-correlation physics.  For all previously
proposed optical flux lattices\cite{ofl,cooperdalibard,Jimenez} the
magnetic flux density is very non-uniform, vanishing at a set of
points.  This non-uniformity is a mathematical necessity for
$N=2$\cite{ofl} and for general $N$ when $\hat{V}$ consists of the
generators of SU(2), as in Ref.~\onlinecite{cooperdalibard}.  The
above considerations show how to overcome this limitation: one should
form a reciprocal-space tight-binding model for which the lowest
energy band has uniform Berry curvature.

\begin{figure}
\vskip0.3cm
\includegraphics[width=1.05\columnwidth]{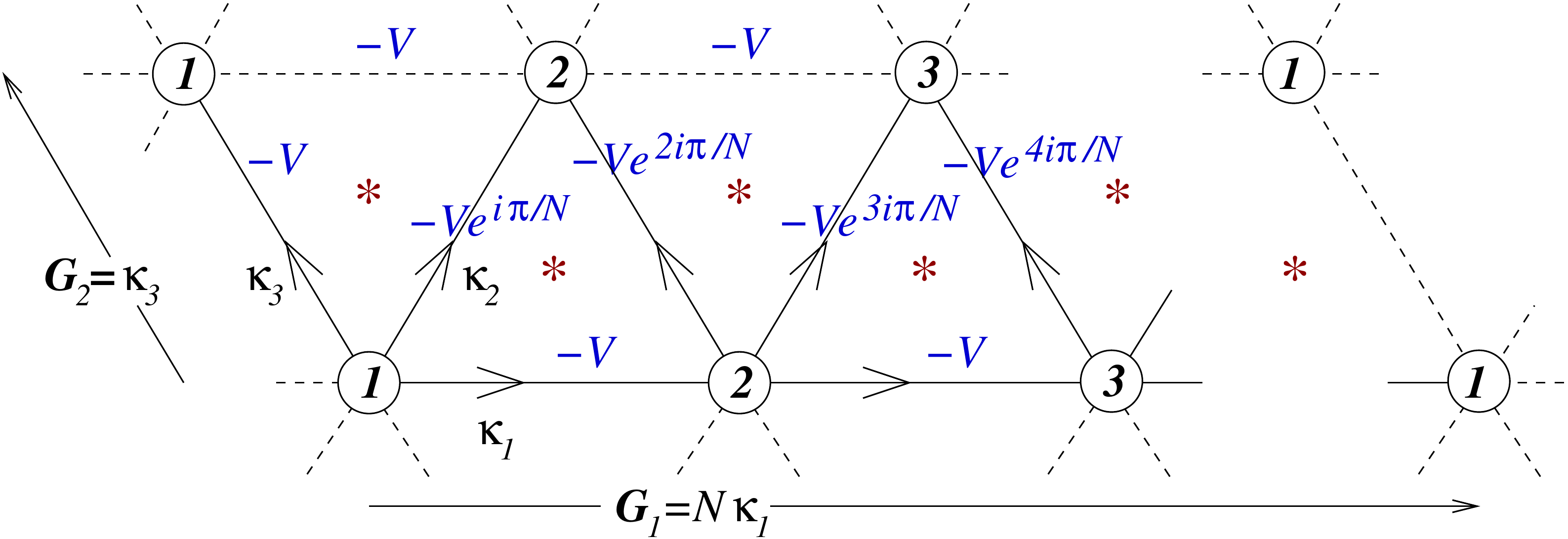}
\caption{
\label{fig:triangularN}
One primitive unit cell of the reciprocal-space tight-binding model
corresponding to the triangular lattice optical flux lattice with $N$
internal states (\ref{eq:vn}), showing the optical couplings $-{\cal
  V} e^{i\phi}$.  The clockwise sum of the phases $\phi$ around any
one of the triangular plaquettes is $\pi/N$ modulo $2\pi$.}
\end{figure}

An optical flux lattice that achieves this goal is illustrated in
Fig.~\ref{fig:triangularN}.  The $N$ internal states are arranged on
the sites of a triangular lattice, marked by the state label
$\alpha$. The links between lattice sites indicate optical couplings,
the displacement in reciprocal space the momentum transfer, with the
label marking amplitude and phase.
Explicitly, Fig.~\ref{fig:triangularN} encodes the coupling matrix
\begin{widetext}
\be
{\hat{V}({\bm r}) = 
-  {\cal V}
\left(\begin{array}{ccccc}
2\cos({\bm r}\cdot {\bm \kappa}_3) & A_1+A_2e^{-i\frac{\pi}{N}} & 0 & \ldots & A_1^*+A_2^*e^{i\frac{\pi(2N-1)}{N}} \\
 A_1^*+A_2^*e^{i\frac{\pi}{N}} & 2\cos({\bm r}\cdot {\bm \kappa}_3 -\frac{2\pi}{N}) &  A_1+A_2e^{-i\frac{3\pi}{N}} & \ldots & 0\\
 0 &  A_1^*+A_2^*e^{i\frac{3\pi}{N}} & 2\cos({\bm r}\cdot {\bm \kappa}_3-\frac{4\pi}{N})  & \ldots & 0\\
\vdots & \vdots & \vdots & \ddots & \vdots \\
 A_1+A_2e^{-i\frac{\pi(2N-1)}{N}} & 0 & 0 & \ldots & 
 2\cos({\bm r}\cdot {\bm \kappa}_3 -\frac{2\pi(N-1)}{N}) 
\end{array}\right)}
\label{eq:vn}
\ee
\end{widetext}
where $A_j \equiv \exp(-i{\bm r}\cdot{\bm \kappa}_j)$, ${\bm
  \kappa}_1= (1,0)\kappa\, , {\bm \kappa}_2=
\left(\frac{1}{2},\frac{\sqrt{3}}{2}\right)\kappa$ and ${\bm
  \kappa}_3={\bm \kappa}_1-{\bm \kappa}_2$.
  We choose the primitive unit
cell of the reciprocal lattice to have basis vectors ${\bm G}_1= N{\bm \kappa}_1$
and ${\bm G}_2= {\bm \kappa}_3$; the real-space unit
cell then has lattice vectors ${\bm a}_1 = \frac{4\pi}{\sqrt{3}\kappa
  N}(\sqrt{3}/2,1/2)$ and ${\bm a}_2 =
\frac{4\pi}{\sqrt{3}\kappa}(0,1)$.

\begin{figure}
\includegraphics[width=0.7\columnwidth]{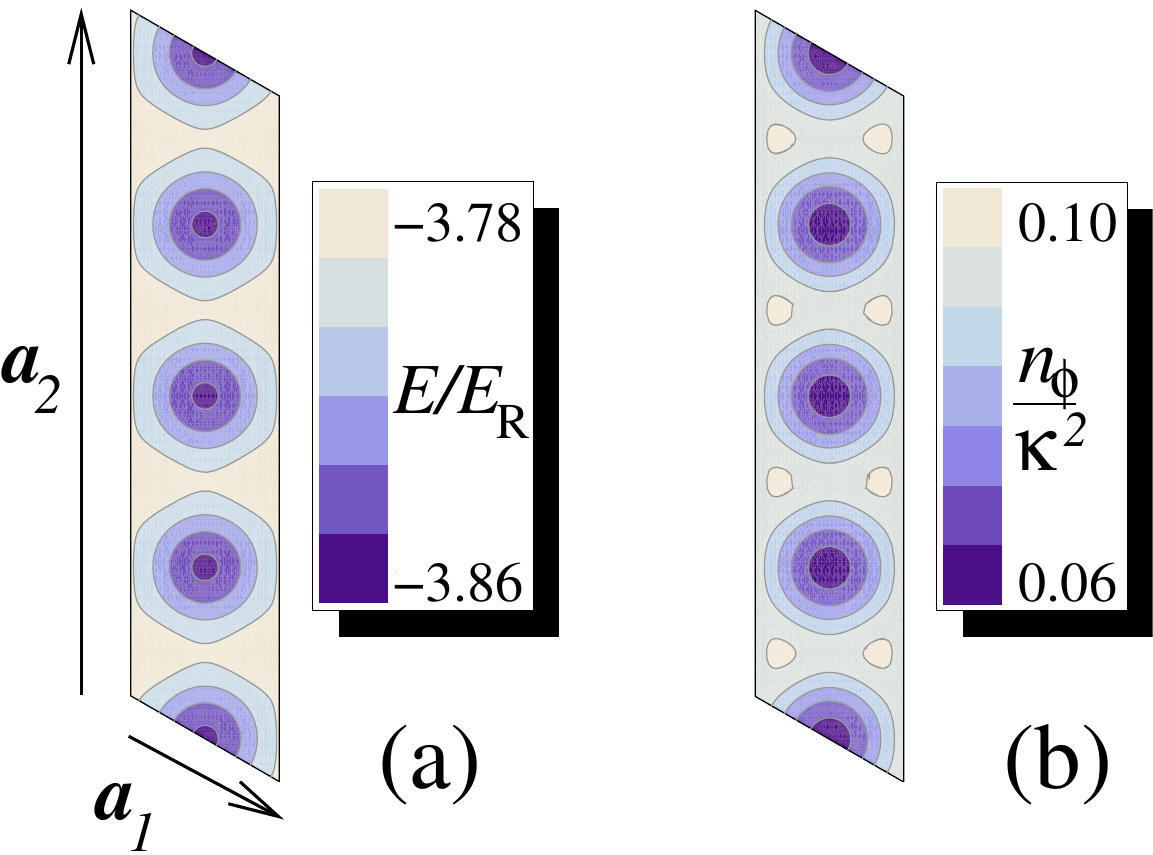}
\caption{\label{fig:spatialn4} Spatial dependence of (a) the energy
  and (b) the magnetic flux density of the lowest energy dressed state
  of the triangular optical flux lattice (\ref{eq:vn}) for $N=4$,
  within a unit cell containing $N_\phi=1$ flux quantum. Both are
  highly uniform, becoming increasingly so as $N\to \infty$.}
\end{figure}

The properties of the resulting dressed states follow from the
bandstructure of the reciprocal-space tight-binding model.  The phases
on the nearest-neighbour couplings mimic the effect of a uniform
magnetic field in reciprocal space (the total phase acquired on
hopping around any triangular plaquette being $\pi/N$, modulo $2\pi$),
so the spectrum follows from Harper's equation for the triangular
lattice\cite{clarowannier}.  For $N\geq 2$ the lowest energy band has
Chern number ${\cal C}=1$, implying an optical flux lattice with
$N_\phi=1$ flux quantum per unit cell, {\it i.e.}  mean flux density
\be
\bar{n}_\phi  = \frac{N_\phi }{         |{\bm a}_1\times {\bm a}_2| } = \frac{N\sqrt{3}\kappa^2}{8\pi^2}\,.
\label{eq:nphibar}
\ee
The local magnetic flux density and the dressed state energy follow
from the Berry curvature and dispersion of the lowest energy band of
the reciprocal-space tight-binding model.
Both show spatial modulations, illustrated in Fig.~\ref{fig:spatialn4}
for $N=4$.  (The spatial dependence of the dressed state takes a form
akin to a triangular multi-component Skyrmion
lattice\cite{kovrizhindoucotmoessner} with lattice constant $|{\bm
  a}_1|$.)  However, as $N\to\infty$, the reciprocal-space
tight-binding model recovers the continuum limit, and its lowest band
becomes a degenerate lowest Landau level with uniform Berry curvature.
The convergence to uniformity with increasing $N$ is very fast, with
corrections that fall exponentially
quickly\cite{clarowannier,kovrizhindoucotmoessner}.

\begin{figure}
\includegraphics[width=0.9\columnwidth]{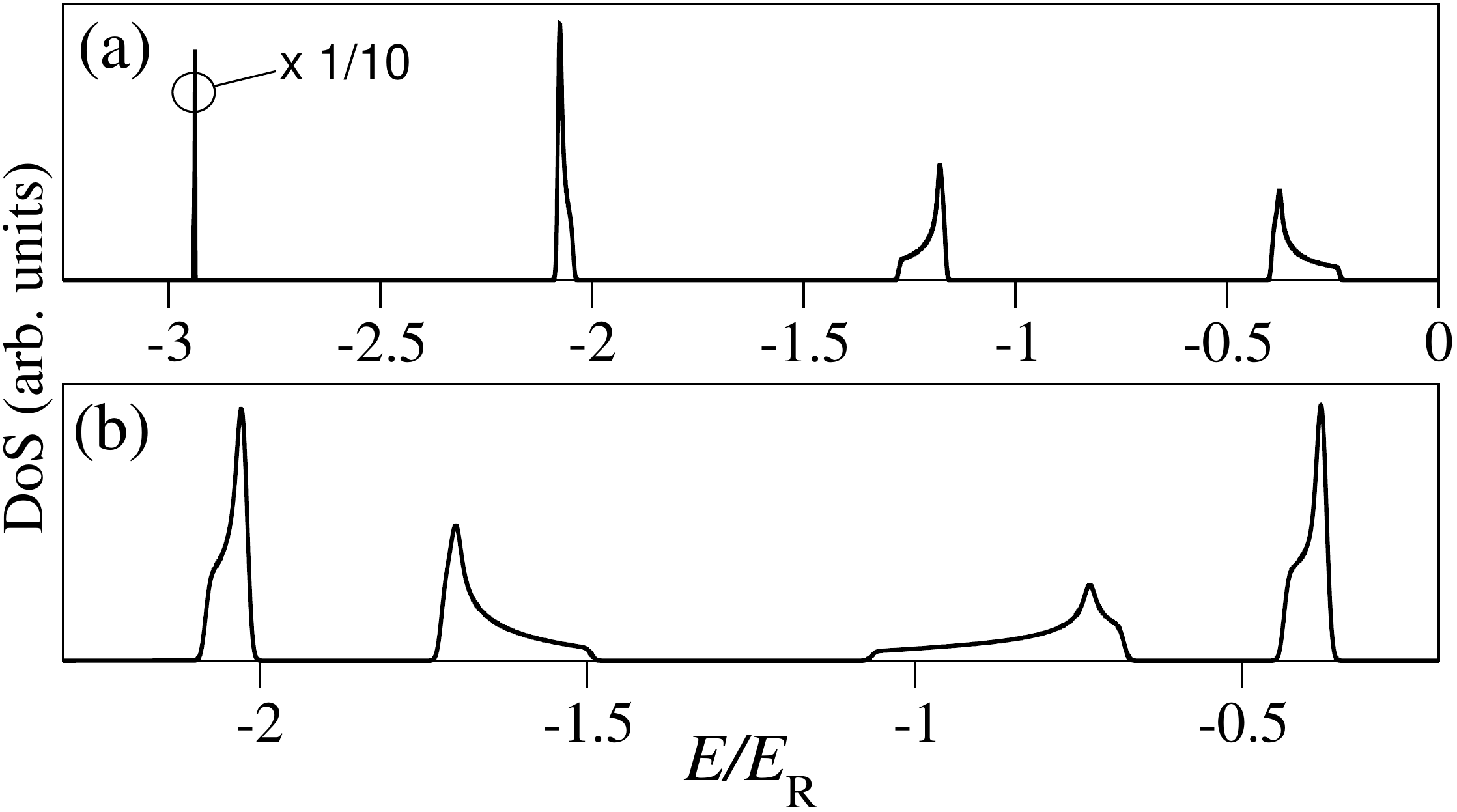}
\caption{\label{fig:spectrum4} Low energy density of states for: (a)
  The triangular optical flux lattice (\ref{eq:vn}) with $N=4$
  internal states and ${\cal V} = E_{\rm R}$; the spectrum closely
  reproduces the Landau level spectrum, all bands shown having Chern
  number ${\cal C}=1$.  (b) A related model (still with $N=4$ and
  ${\cal V} = E_{\rm R}$) in which the phases on the optical couplings
  are modified as described in the text to form a lowest energy band
  with Chern number ${\cal C}=2$.}
\end{figure}

The emergence, for large $N$, of highly uniform magnetic flux density
distinguishes this optical flux lattice from all previous proposals.
Owing to this uniformity, the resulting bandstructure for non-zero
kinetic energy, $E_{\rm R}\ll {\cal V}$, closely reproduces the
spectrum of a charged particle in a uniform magnetic field: degenerate
Landau levels, spaced by the cyclotron energy $\hbar\omega_c =
\frac{\bar{n}_\phi h}{M} = \frac{N\sqrt{3}}{2\pi} E_{\rm R}$.  The
fast convergence means that this qualitative structure is already
apparent for small $N$. This is clearly seen in
Fig.~\ref{fig:spectrum4}(a) for $N=4$, for which the lowest energy
band has a bandwidth that is about $200$ times smaller than the gap to
the next band. (For $N=3$ the same ratio is about $40$.)

One can envisage a variety of experimental implementations for
lattices of this type, or the many variants that achieve the same end.
The case $N=2$ is the triangular optical flux lattice described in
Ref.~\onlinecite{ofl}. For $N=3$ one can use the three components of
the $F=1$ hyperfine manifold of $^{87}$Rb\cite{dalibardprivate}.  A
simple variant is obtained by omitting the state-dependent coupling
along ${\bm \kappa}_3$ (the geometry can then be made square by
rotating ${\bm \kappa}_2 \to (0,1)\kappa$). This lattice requires only
cyclic coupling over the $N \to N\pm 1$ states.  An implementation of
couplings of this kind has been proposed\cite{spielmancyclic} for
$N=4$ using hyperfine levels of $^{87}$Rb.

{\unue{Weak-lattice limit}} Further useful results can be obtained by considering the eigenvalue
problem (\ref{eq:tb}) in the ``nearly-free-electron'' limit ${\cal V} \ll E_{\rm R}$. 
Here, the only influence of the couplings
$V^{\alpha'\alpha}_{{\bm \kappa}}$ is at values of crystal momentum
${\bm q}$ close to the lines where two free-particle states are
degenerate, where 
they open up band-gaps.  Close to those
points ${\bm q}^*$ where two (or more) lines cross, the bands can also
acquire non-zero Berry curvature. Therefore, one can determine
the Chern numbers of the bands in the weak-lattice limit by computing
the Berry curvature in the vicinity of these high symmetry points
${\bm q}^*$.

To illustrate the construction, consider the triangular optical flux
lattice, Fig.~\ref{fig:triangularN}, in this limit. The high symmetry
points, ${\bm q}^*$, are located at the centres of the triangular
plaquettes, at which the kinetic energies of three states (labelled
$a,b,c$) are equal. (These points are marked by $*$'s in
Fig.~\ref{fig:triangularN}.)  This degeneracy is split by the optical
couplings around this plaquette, which we denote $V_{ab} = - {\cal V}
e^{i\phi_{ab}}$, $V_{bc} = - {\cal V} e^{i\phi_{bc}}$, $V_{ca} = -
{\cal V} e^{i\phi_{ca}}$. Integrating the Berry curvature for the
lowest energy state over the vicinity of ${\bm q}^*$ shows that the
total contribution is $\phi_{\rm tot} =
\mod(\pi+\phi_{ab}+\phi_{bc}+\phi_{ca},2\pi)-\pi$.  (For $|\phi_{\rm
  tot}|=\pi$ the lowest two bands cross at a Dirac point, so the Berry
curvature of the lowest energy band is not defined.)  For the
triangular lattice model (\ref{eq:vn}) the phases are such that
$\phi_{\rm tot} = \pi/N$ for each of the $2N$ plaquettes. Thus we
immediately establish that the Chern number of the lowest energy band
is $\frac{1}{2\pi} \times (2N)\times (\pi/N) =1$.  

Our approach allows us to design other forms of optical lattice, with
topological bands that are not directly related to Landau levels.
First we show that one can form optical lattices which have bands with
non-zero Chern number, but for which the adiabatic dressed state in
real space feels vanishing net flux.  In the solid state literature,
these are referred to as Chern bands.  All previous examples involve
tight-binding models in real space, following the seminal work of
Haldane\cite{haldanehoneycomb}.  The analogous ``optical Chern
lattices'' that we describe here have the new feature that the
magnetic flux density is defined as a {\it continuous} function of
real-space position.  A simple example of such an optical Chern
lattice is obtained by choosing the triangular optical flux lattice,
Eqn.~(\ref{eq:vn}), with $N=2$ and introducing a state-dependent
potential $\delta {\cal V} \hat{\sigma}_z$.  For $\delta {\cal V} >
2{\cal V}$ the lowest energy dressed state has $\langle
\hat{\sigma}_z\rangle < 0$. Hence, the Bloch vector\cite{ofl} cannot
wrap the sphere -- the net flux through any unit cell must
vanish. However, this additional term does not affect the Berry
curvatures at the (four) symmetry points in the Brillouin zone in the
weak-lattice limit: the Chern number of the lowest energy band remains
${\cal C}=1$.

Finally, we show how to design an optical lattice for which the lowest
energy band has Chern number of magnitude larger than one, $|{\cal
  C}|>1$, in the weak-lattice limit.
One should arrange that the integrated Berry curvatures from all
symmetry points ${\bm q}^*$ sum to the appropriate value $2\pi {\cal
  C}$. As a concrete example, we adapt the triangular lattice (\ref{eq:vn}) 
 to form a lowest energy band with arbitrary ${\cal C}$ by
replacing the phases $\frac{\pi}{N}$ appearing in (\ref{eq:vn}), in
both diagonal and off-diagonal entries, by ${\cal
  C}\frac{\pi}{N}$. The requirement that the integrated Berry
curvature per plaquette lies in the range $-\pi < {\cal
  C}\frac{\pi}{N} < \pi$, leads to $|{\cal C}| < N$. Hence, 
   a lowest energy band with ${\cal C}=2$ can be achieved for $N\geq
3$ internal states.  Numerical studies for $N=3,4$ confirm that the
lowest energy band retains ${\cal C}=2$ for ${\cal V} \sim E_R$, and
even into the deep-lattice limit. The low
energy bandstructure for $N=4$ for ${\cal V} = E_R$ is shown in
Fig.~\ref{fig:spectrum4}(b). The lowest energy band has Chern number
${\cal C}=2$.  The band is well separated from higher bands, the gap
to the next band being about $4.5$ times its bandwidth, despite the
fact that no optimization has been applied.

The methods we have described open the door for the construction of
many other forms of optical flux lattice. Any tight-binding model for
which the lowest energy band has non-zero Chern number can be used as
the basis for an optical flux lattice.  Hence uniform magnetic flux
density in real space could also be achieved for a system with fixed
$N$ by introducing further-neighbour couplings on the reciprocal-space
lattice\cite{kapitmueller}.  Such couplings are readily implemented as
higher momentum transfers; indeed, the optical flux lattice proposed
in Ref.~\onlinecite{cooperdalibard} involves all the couplings of the
Haldane model\cite{haldanehoneycomb}.  We have also provided a direct
method by which to construct optical lattices with low energy Chern
bands of any topology in the weak-lattice limit.
This ability to design topological bandstructures, with narrow
bandwidth, will allow the use of atomic gases to explore novel
correlated topological phases of bosons or fermions in lattice
systems.

\vskip0.1cm

\acknowledgments{NRC is grateful to Jean Dalibard for numerous 
  helpful discussions, and RM to Beno\^it Dou\c{c}ot and Dima Kovrizhin for collaboration on Ref.~\cite{kovrizhindoucotmoessner}. 
  This work was supported by EPSRC Grant
  EP/F032773/1. We thank the Galileo Galilei Institute for
  hospitality.}

\end{document}